# Teaching Machine Learning in K–12 Computing Education: Potential and Pitfalls

**MATTI TEDRE[1], TAPANI TOIVONEN[1], JUHO KAHILA[2], HENRIIKKA VARTIAINEN[2], TEEMU VALTONEN[2], ILKKA JORMANAINEN[1], AND ARNOLD PEARS[3] (Senior Member, IEEE).**

[1]School of Computing, University of Eastern Finland, PO Box 111, 80101, Joensuu, Finland (e-mail: firstname.lastname@uef.fi)
[2]School of Applied Educational Science and Teacher Education, University of Eastern Finland, PO Box 111, 80101, Joensuu, Finland (e-mail: firstname.lastname@uef.fi)
[3]School of Industrial Engineering and Management, KTH Royal Institute of Technology (email: pears@kth.se)

Corresponding author: Matti Tedre (e-mail: firstname.lastname@uef.fi).

**ABSTRACT** Over the past decades, numerous practical applications of machine learning techniques have shown the potential of data-driven approaches in a large number of computing fields. Machine learning is increasingly included in computing curricula in higher education, and a quickly growing number of initiatives are expanding it in K–12 computing education, too.

As machine learning enters K–12 computing education, understanding how intuition and agency in the context of such systems is developed becomes a key research area. But as schools and teachers are already struggling with integrating traditional computational thinking and traditional artificial intelligence into school curricula, understanding the challenges behind teaching machine learning in K–12 is an even more daunting challenge for computing education research. Despite the central position of machine learning in the field of modern computing, the computing education research body of literature contains remarkably few studies of how people learn to train, test, improve, and deploy machine learning systems. This is especially true of the K–12 curriculum space.

This article charts the emerging trajectories in educational practice, theory, and technology related to teaching machine learning in K–12 education. The article situates the existing work in the context of computing education in general, and describes some differences that K–12 computing educators should take into account when facing this challenge. The article focuses on key aspects of the paradigm shift that will be required in order to successfully integrate machine learning into the broader K–12 computing curricula. A crucial step is abandoning the belief that rule-based "traditional" programming is a central aspect and building block in developing next generation computational thinking.

**INDEX TERMS** Machine Learning, Artificial Intelligence, K-12, School, Computing Education, Computational Thinking, Computing Education Research

## I. INTRODUCTION

COMPUTER-BASED automation of jobs has driven changes in the labor markets since the stored-program computer revolution started to gain momentum in the 1950s. The first jobs to disappear were routine tasks of a symbolic or numerical nature that were amenable to computing-based automation through explicitly stated sets of rules. Examples are many, but some notable ones are accounting, payroll, inventory, and scheduling, for example [23]. Governments used computers for large-scale information processing needs, such as compiling and tabulating national census data and welfare state record-keeping [1]. Scientists gradually adopted the new technology for tasks that involved large-scale computations, such as X-ray crystallography and fluid dynamics [1], [2]. Ever-increasing volumes of digitized data and increasing processing power drove the rise of scientific computing, culminating during the 1980s in computational sciences movements in multiple fields and the emergence of e-science on national political agendas [12].

Pioneering educators quickly recognized the changes in the labor market. They expressed very early the need to educate everyone about the new technology that they saw was bound to change work and society [25]. Guzdial [31] wrote that throughout the years, advocates of computing and programming education in K–12 have provided many rationales for their work. Proponents argued that knowing the principles of computing was important to understanding the algorithm-driven virtual and physical worlds one inhabits. In addition, familiarity with these concepts enables people to use computers better and to ask questions about the powerful influence of algorithms on their lives. They have also argued that computing provides a new medium for expressing ideas and for studying and learning science, mathematics, problem-solving [64], and processes in general [31]. As computing pervades work life it becomes more important, emerging in recent years as a central job skill and an area closely connected with economic success and industrial innovation [29].

The abstraction level in programming (or, more broadly, in controlling computations) has steadily risen, with its cited benefits ranging from improved business efficiency to bringing the power of computing from hands of the few to the hands of many [18], [26], [27], [50], [63]. After the 1950s, interfaces to computing power have been rapidly moving further and further away from the machine [28], and decade after decade new educational interventions have aimed at capturing each era's dominant approaches to automation [96].

As the level of abstraction in computing education across educational levels steadily arose, the credo among computing cognoscenti became that one needs to be familiar with at least one abstraction level below that at which one is working. In the 1950s it was said that computing personnel benefit from knowledge of the underlying electronics [40]. Then it was said that knowledge of octal machine code was beneficial for those who programmed in assembly languages [39]. Once a consensus had formed on the adoption of higher-level languages such as Pascal, C, C++ and Java, knowledge of assembly language was correspondingly identified as important. Knowledge of how to implement data structures and algorithms is beneficial for users of highly optimized class libraries. This article concerns another step in this development, and it asks what role does an understanding of high-level language "traditional" programming play in the adoption of AI and ML computational toolboxes and languages.

The scope and focus of computing education efforts have always expanded and shifted to keep pace with technological change (cf. [31], [32], [96]). Since the birth of modern computing, year by year educational interventions expanded to involve ever younger and more diverse groups of students; in the 1960s they reached first high schools and then primary schools and kindergartens. In 1960 Alan Perlis argued that soon everyone needs to learn programming and "algorithmizing," but lamented the lack of pedagogy needed in order to effectively teach this [50]. The mid-1960s brought DEC PDP-8 minicomputers to select US high schools, the 1965 Little Man Computer initiative taught machine languages to children, the children's programming language Logo saw daylight in 1967, and the children's portable computer Dynabook was introduced in 1968 [14].

In the past ten years the computing landscape has seen another major technological shift. Traditional programming and rule-based "good old-fashioned artificial intelligence", which have been the driving force of automation for the past 70 years, have been joined by a variety of data-driven machine learning techniques. The much-hyped "second machine age" [9] is based on the ability of machine learning techniques to automate many tasks that traditional, rule-based programming struggles with. In many application areas it has turned out that for large classes of problems it is much easier to collect data sets large enough for machine learning than to figure out the rules necessary for a rule-based program [49]. Many popular examples of the latest advances in automation are based on a combination of neural networks and traditional programming: take, for example, self-driving cars, face recognition [90], computer based identification of tumors [20], and the game of Go, where a computer was programmed to learn completely on its own to achieve superhuman ability in the game [83]. Reports of the recent, much feared, losses of jobs to automation in the knowledge-work domain cite numerous examples where job losses were principally not linked to traditional rule-based program development, but rather to areas where the structure of the tasks have been amenable to the sophisticated optimization and statistical techniques of modern machine learning [11].

Just as previous developments in computing have triggered changes in computing education, machine learning is now acting as a catalyst for change throughout the education system both in K–12 and higher education. The focus of computing education has shifted before, new shifts will come in the future, and is aligned with suggestions that the next frontier in computer science education research is how to teach artificial intelligence [15], [81], [82].

Computing pioneers have foreseen a shift in higher education computing curricula towards machine learning principles [82]. What is more, the breakthrough of machine learning has given rise to a growing number of initiatives for teaching some machine learning principles in K–12 [59], [99]. A 2020 study identified 30 recent educational initiatives that focused on ML basics and neural networks in K–12 education [59]. Some pioneering ones include the Wolfram Alpha-based "Machine Learning for Middle Schoolers"[1], Google's Teachable Machine[2], and the IBM Watson-based "Machine Learning for Kids"[3]. Examples of research on bringing machine learning to children at different ages include combining block-based languages and APIs with popular cloud-based machine learning systems [47], extending block-based environments with ML blocks (using, e.g.,

---

[1] https://writings.stephenwolfram.com/2017/05/machine-learning-for-middle-schoolers/
[2] https://teachablemachine.withgoogle.com/
[3] https://machinelearningforkids.co.uk/

Cognimates[4] and eCraft2Learn[5]) [41], [46], [92], [118], AI-enhanced children's robotics kits (e.g., Calypso for Cozmo[6], AI-in-a-box[7]), and teaching the principles of object recognition to high-school students [58]—along with numerous others [59]. Public efforts are under way by, for instance, the AI4K12 working group for teaching AI through kindergarten, primary school, and high school [99].

While the K–12 machine learning initiatives are of many kinds, most of them address one of the chief rationales for computing education for everyone [31]: The importance of understanding how one's world works. They are aimed at bridging the gulf between K–12 computing education and children's everyday experiences with technology. Without understanding some ML principles, many apps and services that children use appear like magic to them: Picture hosting services know who the people in one's photos are, streaming websites are able to recommend videos that the user will like, even though he or she has never seen it before. Mobile phones can be configured to unlock when they see their owner's face, and home assistants act on voice commands. Understanding that all the above are smartly designed technology, which is in no way intelligent in the same way humans are, is important to demystifying these technologies [37], [41].

ML in K–12 initiatives also prepare students for citizenship in a new media ecology driven by ubiquitous data collection, profiling, and behavioural engineering [55], [65]. With suitable curricular and educational arrangements, learning the principles of ML can provide an antidote to malicious data collection, attempts to sway elections, and profiled advertisement [103]. Understanding what ML can, and cannot, do is an important enabler of active citizenship and underpins the future prosperity of democratic societies [34]. Enabling people to ask questions about the powerful influence of the most famous, as well as the most notorious, algorithms on their lives—those that enable tracking, profiling, modeling, predicting, content tailoring, and behavior engineering, for instance—requires a perspective different from that associated with traditional rule-based programming. Calls for education that prepares students for the data-driven society in which they live are becoming frequent [16], [68], [103].

But teaching ML in K–12 differs in a number of ways from the traditional approaches to teaching rule-based programming, computational thinking, or computing. This article is aimed at explaining why there is a growing belief that how to teach ML is going to be the next frontier in computing education research [15], [81], [82]. The article describes a number of pedagogical and technical aspects of computing education that ML learning interventions in K–12 have had to re-think, as well as their implications to educational practice. The article is aimed at K–12 computing educators who are planning to expand their educational portfolio towards ML-based technology. While ML in K–12 initiatives are not all alike, many of them share a number of features identified in this article, and understanding these approaches provides an important basis for future educational innovation.

## II. METHODOLOGY
Other reviews of machine learning in K–12 have found that traditional systematic reviews on the topic are not well suited for the task due to 1) the novelty of the topic, 2) the lack of established terminology, and 3) broad variety of disciplinary approaches in existing peer reviewed literature on AI education (e.g., [54], [117]). Those studies have found that exploratory literature reviews fare much better in reaching literature relevant to the topic than structured literature reviews do [54], [117]. Similar to the earlier reviews, this study too adopted the scoping studies framework, which aims at rapidly mapping an emerging field, including, for instance, its key concepts and their relationships, its key findings, and its research gaps [6]. Scoping reviews are particularly well suited for mapping a terminologically eclectic field like ML education in K–12 because, unlike traditional systematic reviews, the scoping review framework does not rely on a fixed, pre-defined set of search terms for identifying relevant research [6]. The weaknesses of scoping reviews include their inability to appraise quality of identified research and quantitatively synthesize results (as compared to meta-reviews), as well as their poorer replicability [6]. Instead, they are able to provide a narrative account of available research and summarize it [6].

In order to identify the literature on the topic, this study began by conducting a set of searches in ACM Digital Library, IEEExplore, Scopus, and Google Scholar, using sets of keywords borrowed from [54] and [59]. From the result sets—the size of which ranged from zero to 62.200—a maximum of 150 abstracts were scanned from each, identifying research focused on ML at K–12 education. Second, in order to extend the overview of the field, snowball sampling was applied to the bibliography sections of the articles already identified [111]. Third, in order to raise the analysis to a higher level of abstraction, similar studies were clustered in order to understand the groundings for the new emerging research area. The results found 63 articles relevant to learning ML in K–12. The size of the resulting set differed from the other reviews due to their different foci (33 instructional units in [59], 150 AI-education related documents in [54], and 49 K–12 AI education works in [117]). As the field is still emerging and growing, and as its terminology is not yet fixed, the set of documents captured may not be exhaustive.

From the identified 63 documents, education researchers and computing researchers distilled their lists of key characteristics of K–12 ML education (in contrast to traditional computing education), yielding an original working list of 19 characteristics. Those key characteristics were discussed, sorted, and merged to a final list of 13 items (Section IV). Each category was described through a number of exemplars, key articles that best exemplified them.

---

[4]https://cognimates.me/
[5]https://ecraft2learn.github.io/ai/
[6]https://calypso.software/
[7]https://www.readyai.org/readyai-you/ai-in-a-box/

## III. MACHINE LEARNING EDUCATION IN K–12

**T**HEORY AND PRACTICE of artificial intelligence (AI) have belonged to the fundamentals of higher education in computing since the birth of the AI field in 1956 [60], [74]. In higher education, much education on the theory of AI today focuses on the mathematical foundations of algorithms for building models that are able to generalize and predict from unstructured data. And much education on AI practice aims to harvest the power of these algorithms by focusing on applying and using AI tools, models, and algorithms rather than rigorously explaining the underlying structures of the algorithms and the learning of the algorithms themselves.

In K–12 education, most AI-related initiatives have historically been concerned with 1) AI-based tools to support learning, 2) AI-based tools for studying learning processes, and 3) AI to support school administrative functions [38]. Adaptive learning environments, pedagogic agents, automated governance, intelligent tutoring systems, and many other similar research programs have attempted to model elements of the learning situation—such as the learner, pedagogy, subject matter, context of learning, and learning objects—in ways amenable to automation [5], [38], [110]. As technology and educational paradigms have changed, so have views about the merger between AI and education [5].

In addition to using AI in educational technology, over the decades there has been an important undercurrent: (How) can we teach children about how AI works? Since the 1970s AI education initiatives in K–12 have followed each era's "hot topics in AI" with examples spanning initiatives with children programming robots to navigate in the world [66], children working with NLP (natural language processing) technology [45], children developing expert systems in the classroom [106], [107], middle-school students teaching a computer to play tic-tac-toe [24], [69], all the way to high school students learning the fundamentals of neural networks [7]. LEGO Mindstorms has been a very popular platform for robotics-based AI education in K–12 [67], and AI initiatives in schools have very often been based on developing rule-based systems.

In higher education the symbolic, rule-based "classical" AI has been joined by data-driven branches of AI, today most commonly machine learning (ML). Educational efforts have followed suit [114]. Major computing publications have outlined the impact of machine learning on the undergraduate computing curriculum [81], [82]. For example, Shapiro et al. [82] define the traditional view of the core of computing as a collection of human-comprehensible, deterministic algorithms that can be verified. They envision two shifts away from this in the near future. They observe that machine learning models are not human-readable algorithms but opaque composites of millions of parameters. Secondly, ML models are not amenable to verification efforts, however, their effectiveness can be statistically established. They further note that nearly all literature on computing education research targets rule-driven programming, and consequently call for a major shift in the focus of computing education research to study how people learn and reason about ML systems [81], [82].

Multiple modern programming languages now offer suites to build and train ML models without exposing the underlying operations and the architecture of ML solutions. Languages such as Python, R, and Matlab have been used in practical and theoretical courses on ML education in tertiary education. The trend in ML education is towards increasingly sophisticated tools for enabling practical hands on experiences for students. The interest in deep learning algorithms in higher education is no longer just about learning the equations of back-propagation or ReLU, or how to implement them; but about applying them as tools that can be used with little knowledge of their internal structure and mechanisms. Through easy-to-access APIs ML becomes a commodity and agency has shifted from the hands of experts to the hands of undergraduate and graduate students. What is more, unplugged activities can be used to scaffold understanding of the training processes used on data sets to produce classifiers in a range of realistic settings [52].

Today, as ML-based applications have become a common part of children's everyday life, ranging from smart toys to music streaming services, attention has turned to how to teach some ML principles to children [99], [114]. In order to render ML accessible and democratize the access to these technologies the sophisticated underlying model, and details of its internal implementation, have been buried under layers of abstraction. Due to the complexity of many ML algorithms, and due to the black box nature of ML's predictive models, theoretical and practical ML education use a broad spectrum of tools to soften the learning curve for state-of-the-art ML solutions. For instance, some projects have used various sets of scripting and notation languages to scaffold the black boxed parts of the predictive models [42], [43]. Other projects have used mobile robots to introduce the concepts of artificial neural networks (ANNs) and selected ML paradigms, such as reinforcement learning [97], [98]. Many of these projects use mobile robots in ML education to transform a theoretical subject into a tangible, practical and explicit representation of the predictive models.

However, like other branches of computational thinking, one key challenge of ML in K–12 education is determining what can be taught at different levels of education [14]. The phrase "K–12" covers a broad range of learners and different age-appropriate learning objectives, spanning from kindergarten children who focus on learning basic literacy and numeracy to senior year high school students, many of whom are working on mathematical modelling and advanced problem solving skills in order to enter tertiary education. The tools and pedagogical approaches appropriate to each age, learning context, and educational objective are very different, and the spectrum of challenges and aims of ML education initiatives reflect these great differences. ML education initiatives in K–12 are of extremely many kinds, and often in addition to ML concepts and skills also involve learning about fundamental principles of time and place, as well as understanding and being able to manipulate materials, constructions (physical artefacts), as well as being able to place such artefacts into a broader historical perspective.

A large number of initiatives have emerged in the context of K–12 education to fill the need to teach middle-school children some basic ML concepts. In addition to the above mentioned *Machine Learning for Middle Schoolers* based on Wolfram Programming Lab, Google's *Teachable Machine 2*, and the IBM Watson-based *Machine Learning for Kids*, a number of other initiatives have approached ML education in K–12 from different angles. Pre-K and kindergarten AI curriculum has been approached using a social robot based on Lego blocks, motors, sensors, and smartphones [108], [109]. Facial recognition and robotics has also been used to teach AI concepts [37], and experiments have been made with gesture recognition in order to study how children learn to understand basic ML workflows [35]. Block language extensions have been adopted to teach high school children concepts like *k*-means clustering, neural networks [21], supervised learning [113], LSTM (long short-term memory) based language models [105], and APIs that connect to commercial ML systems [47]. There are examples of teaching the principles of object recognition to high-school students through a simplified ML system [Authors, 2019a] and studies of children's interactions with different AI-based tools— Jibo robot, Anki's Cozmo robot, and Amazon's Alexa Echo, and Cognimates (conversational agents)—to study how children's understanding of AI technology developed through interactions with smart agents [15], [16].

A number of ML-related initiatives also focus on pedagogical and curricular perspectives. Unified AI literacy curricula from kindergarten to high school and university have been developed [48] as well as separate curricula for different school levels [10], [22], [75], [116]. One initiative presents a 450-hour complete AI curriculum, including 100 hours on machine learning [87]. Extensive packages of teaching materials have been developed to help middle school students to learn about ML [78]. One group presented five "big ideas in AI" that children should learn: Perception through sensors, reasoning about the world through models, learning from examples, interacting with people, and societal impact of AI [99]. Another group described the principles underlying a constructionist AI curriculum for K–9, based on the PopBots robot toolkit, AI ethics, and the Droodle creativity game [3]. A recent study synthesizes AI-related literature to form 17 competencies and 15 design considerations associated with a learner-centered AI curriculum [54].

Some recent ML education systems are very easy to adopt for classroom use. For instance, Google's Teachable Machine 2 (TM2) and MIT's PIC [92]. Abandoning the deductive reasoning and rule-based programming that drove traditional programming language experiments in education, using TM2 or PIC, children can engage in the process of data-driven reasoning and design: providing the machine with a training data-set and then using the trained model to control the machine [Authors, 2020a]. With TM2, children can, for example, examine external representations by having a computer learn to recognize their voice, facial expression, or bodily gestures [99]. These new tools and computational paradigms can expand the action possibilities for children while also offering them new ways to make sense of the world they already live in. Understanding how ML models the world can also empower children to understand and question ML systems of their everyday life, such as those used in face recognition, voice recognition, and other kinds of pattern recognition [37], [120].

## IV. WHAT CHARACTERIZES K–12 ML EDUCATION?

SO-CALLED "LOW-FLOOR" tools [72] are increasingly democratizing and commodifying ML education, and enabling teachers to adopt ML as a basic building block of their computing classes. ML-based tools have become a commonplace element of the every-day user experience. Ever improving interfaces for training ML models have the potential to shift end-user programming towards data-driven applications. In the past few years, a growing body of literature on how to teach ML at different school levels provides concrete indications of where elements of change in school education are taking place. This section presents an overview of some of those changes.

### 1. Broader classes of example applications become accessible for different levels of K–12 education.

Similar to how ML has enabled new approaches to automation of jobs in work life, ML is enabling new kinds of example applications of computing in the school classroom. Take, for instance, media applications, which have been a feature of computing education for decades: There is a long history of research on using sound, images, and movies as a vehicle for learning computing [30]. As machine learning really shines with media-related application areas—video, pictures, sensors, and sound—it has no shortage of immediate, real-world uses in K–12 education. Any phenomenon that allows easy collection of a lot of data has good potential for learning about ML-based technology. That is a strength of machine learning in general: As Karpathy [49] noted, it has turned out that for large sectors of real-world problems, collecting the necessary data for implementing desirable behaviors through ML is much easier than writing a traditional program to do the same. Consequently, many K–12 ML initiatives are media-centered [35], [46], [112] and offer very malleable, accessible tools for a variety of media-related areas, such as describing emotions through music [3], image recognition [58], [112] [Authors2019b], and speech recognition and synthesis [46] [Authors2020a] [Authors2020b].

The ability of K–12 student-made ML projects to easily achieve nontrivial results—the "low floor and high ceiling" principle [72]—has been highlighted as a motivating factor for students [21], [35]. In one series of ML workshops, middle-school children's project ideas included, for example, an app that identifies edible mushrooms and warns about poisonous ones, an app that helps color blind people to identify colors, and an app that recognizes cheerleader poses to help training [Authors2020c]. Figure 1 shows children's interface design for an app that used speaker recognition to report to the teacher which kids make the most noise when the teacher was away [Authors2020a] (that app, ideated by children, also serves as a good example case with which to discuss AI ethics.)

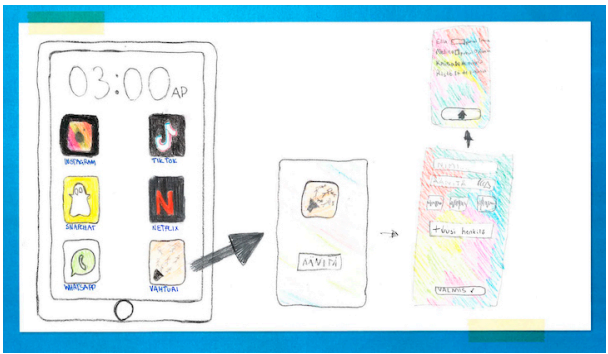

**FIGURE 1.** Sixth-graders' interface design of the voice recognition based "Watchman" app, which they designed for recognizing and reporting to teacher which kids make noise when the teacher leaves the room [Authors2020a].

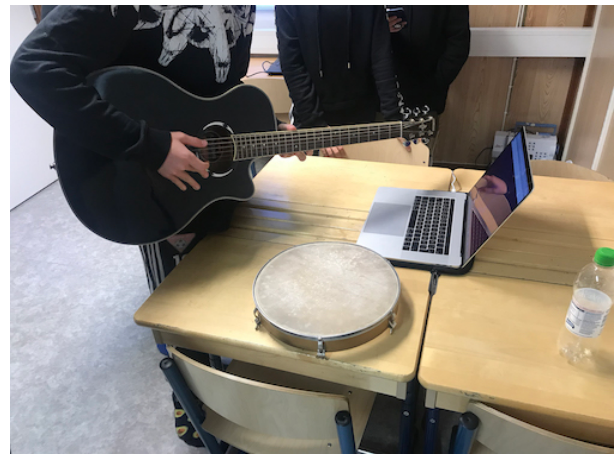

**FIGURE 2.** Ninth-grade students recording samples of musical instruments they wanted their ML app to recognize [Authors].

Pedagogically, co-designing real-world applications that have immediate uses in the children's world is based on the assumption that the iterative process of creating ideas and external representations of them, followed by a re-description and refinement of them, can lead to increasingly sophisticated understandings of the content domain [19], [51]. The pedagogical potential of co-design is based on the fact that during acts of collaborative invention, making the ML models, and testing and co-developing of their functionalities, the students are not only sharing their evolving ideas and understanding, but also creating a complex set of personally meaningful relations between abstract concepts and real-life problem-solving contexts at hand [19]. Accordingly, the pedagogy inherits the idea that understanding domain knowledge can be facilitated by the so-called "21st century skills," such as the ability to collaborate to solve complex problems, to adapt and innovate in response to new demands and changing circumstances, and to use technology to create new ideas and knowledge [77]. The expansion of action possibilities with respect to truly interesting learning tasks is also deeply connected to the perceived ownership of learning and design [73]. K–12 computing initiatives through ages have emphasized creativity; that emphasis is continued with ML [3].

### 2. Focus shifts from rules to data.

Instead of relying on explicitly hand-coded rules that a computer will follow on different inputs, many ML initiatives guide students to train ML models by giving the system a lot of data to learn from [21]. Studies have used drawings, poses, speech, and video [Authors2019b, Authors2020a, Authors2020b], as well as data from, for instance, tracking sports activities [36], [120], webcam [92], [93] gestures [35], web searches [112], and cartoon pictures about kids and mock data about them [88]. As a result, how to curate, create, clean, label, and feed the training data has become a central learning objective for many machine learning initiatives in school computing education [35], [46], [54], [88], [112] [Authors2020c]. The concerns in making those kinds of ML systems work well in the classroom are much more about the quality of data than about choosing the right rules; for instance, one study looked at learners understanding the impact of sample size, sample versatility, and negative examples on ML model [36]. That computers can learn from data has been seen as one of the key lessons of teaching AI in K–12 [22], [75], [99].

Many ML in K–12 initiatives have found that even though the feature extraction methods in popular K–12 ML systems may work reasonably reliably and be capable of capturing the essential information from input, the student-produced data has always noise and unwanted features in it—most likely much more so than data sets produced by domain experts working with traditional ML systems has [Authors2020a, 2020b] [22], [35], [36]. For instance, photos taken by the children are often out of focus, audio input can pick up plenty of unwanted background noise from the classroom [Authors2020a,2019b], and one app that children designed to tell whether the speaker is a boy or a girl was unreliable because primary school boys and girls sound much alike [Authors2020c]. Figure 2 shows an example of one of the more brittle models: Ninth-grade students training an ML model to recognize instruments from each other.

### 3. Emphasis of syntax and semantics changes.

Syntax is one of the many sources of cognitive load in learning programming [53]. In programming education research and psychology of programming, numerous initiatives have addressed the challenges with learning syntax—recently through, for instance, block-based languages [80]. Initiatives on teaching ML in K–12 vary by their take on the roles of syntax and semantics. Some initiatives have shifted the emphasis away from syntactic and semantic concerns, and have allowed young children to engage in exploration of how to control computations without the need to learn a new syntax ( [21], [35] [Authors2019a, 2020a]). Other initiatives, like Wolfram Programming Lab, use traditional programming language syntax to teach students ML topics [112]. Yet others, like Snap!, combine easy-to-approach ML elements and cloud services with block-based programming [46], [47], or enable machine learning within Snap!, like the SnAIp! project does [41]. Several projects have developed ML extensions for the MIT App Inventor [92], [118]. One initiative taught young learners how to create an artificial neural network (ANN) based model by designing the ANN structure—number of inputs and outputs, hidden layers, and units per layer) plus a few simple rules to implement backpropagation [21]. An example of a greatly simplified approach is Google's Teaching Machine 2, which allows the creation and testing of ML models without either classical programming or blocks—however, as an outcome students do not learn concepts central to traditional programming, such as the syntax and semantics related to loop variables, control structures, conditions, or variable types. TM2 requires those important concepts be taught in programming-oriented classes.

### 4. ML allows age-appropriate shifts in pedagogical entry points in classroom education.

In its quest to find working solutions for guiding the formation of problem-solving schemas in computing classes, computing education research has seen a broad variety of pedagogical approaches over the years [85]. Cognitive-load-theory oriented, constructionist, problem/discovery/inquiry-based approaches—along with many others—have inspired a multitude of learning designs for computing education, with no silver bullet in sight yet [31], [85]. ML education initiatives in K–12 have relied on a similarly broad spectrum of pedagogical approaches, with their keywords ranging from project based learning, constructionism [10], [76], creativity [3], experiential learning theory [21], [88], co-design [115], and cognitive apprenticeship [88], to gamification [70] and active learning [10], [22], [87]. Attention has been paid to make education from kindergarten to high school age-appropriate [33] and cumulatively progressive [48]. In kindergarten, for example, ML education might focus on playful exploration and awareness building, in the middle school on experimenting on problem-solving and some theory, and in the high school on fostering core knowledge and exploring advanced topics [48] [Authors2019a, Authors2019b, Authors2020a, Authors2020b].

More often than not, however, reports on ML initiatives discuss technology, content knowledge, techniques, or curricula instead of pedagogical principles and didactic elements of those initiatives [59]. In research reports from ML workshops [Authors2019a] [Authors2020a], teaching the machine to recognize moods, sounds, speech, and poses has resonated well with modern pedagogical approaches that support children's agency through playful and creative learning, bodily interaction, and collaborative advancement of ideas and understanding (e.g., [44], [71], [79]). What is more, many ML based learning environments provide an immediacy of action: As the models are trained, one can right away experiment on them and judge how well they work. Such feedback loop also challenges children to reason about the relationship between their new inputs and the output provided by the interactive ML tools [Authors2020d].

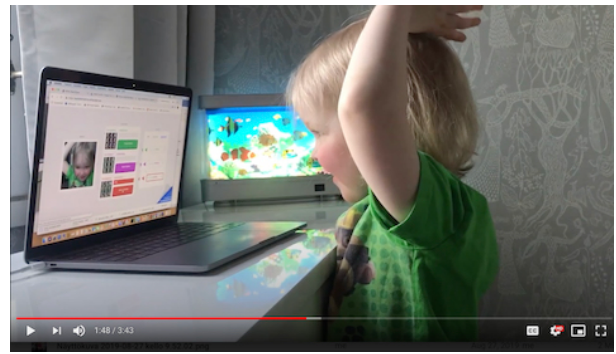

FIGURE 3. A 3-year-old learning to identify and describe his own emotions by teaching a computer to recognize them [Authors2020].

Figure 3 shows an example from a study where very young children learned to identify and describe their own emotions by teaching a computer to recognize them (kindergartens use flash cards for similar purposes). History of learning with educational robotics from Logo turtles to Lego Mindstorms has shown the benefits of tangible design activities in which interaction with people, materials, tools, and technology mediate collaborative construction of ideas [71]. But the ability to control a computer by using one's body to train the computer to recognize things—faces, expressions, poses, objects, and so forth—and then defining actions based on those, promises new approaches to making computing engaging, and it promises a whole horizon of modern pedagogical entry points. Recent research results have emphasized bodily interaction with ML systems [15], [35] [Authors2020c,2020a]. Other research studies have shown the feasibility of teaching ML concepts by incorporating them in athletic practice, and having high-schoolers build, use, evaluate, and iteratively improve machine learning models of athletic skills [120]. These initiatives, along with many others, have emphasized the importance of positioning children and youth as active subjects and teachers of ML systems, and not the objects of teaching typical in more traditional models of instruction in education.

### 5. Interaction with the computer keeps shifting towards more bodily forms and use of natural language.

Throughout the development of computing, the locus and modalities of "user interface" have continuously changed, and that development trend has always been away from the computer and towards the user and communities of users [28]. Continuing the more than 70 years of advances in interface design, learning ML by creating ML models offers new opportunities for bodily interaction and natural language [Authors2019a,2020a] [16], [35], [36], [47], [99]. For example, educational technology for learning ML principles allowed children to train PoseNet-based ML models to recognize different cheerleader poses (Fig. 4) [Authors2019b].

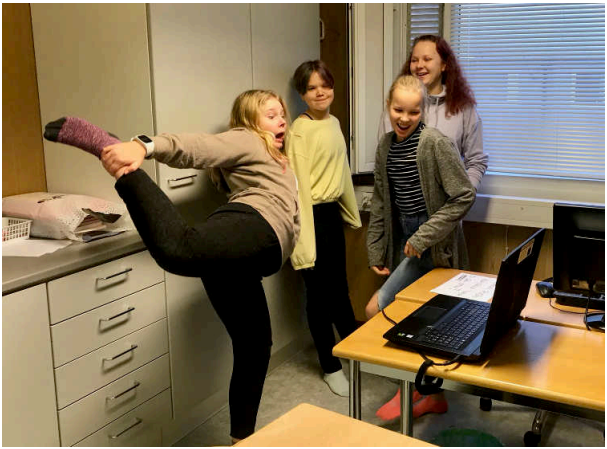

FIGURE 4. Twelve-year old children training a ML model to recognize cheerleader poses [Authors2019b] (GDPR compliant photo).

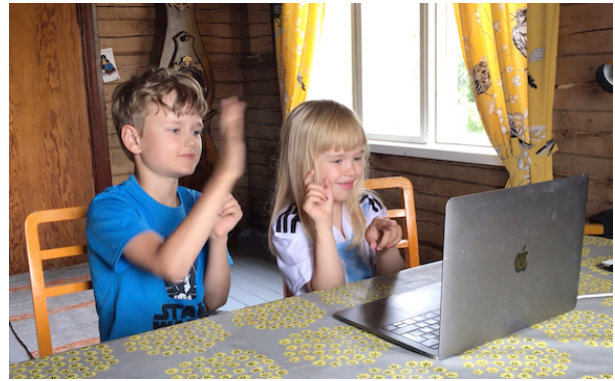

FIGURE 5. A preschooler and older brother studying at home, creating a model to recognize gestures [Authors] (GDPR compliant photo).

The changes in interface modalities are not only related to what pedagogical models does ML match well (see the section above), but also to teaching the principles of ML: One of the "big five" ideas in K–12 AI [99] is that computers perceive the world using sensors. Children's interaction with ML/AI-based tools has been studied extensively, and several studies report anthropomorphic tendencies with very young children, such as viewing robots more as peers than gadgets [91], seeing them as "people" who are less smart than them [108], and consequently, developing relationships with them and seeing them as psychological nonliving things [17], [33], [100]. In ML education, children have felt a two-way relationship where they can teach the machine, instead of just the machine teaching them [Authors2020a] [15].

### 6. The status of algorithmic steps changes.

Much of computing education in schools revolves around the concept of algorithm as it is typically described in computational thinking: step-by-step methods that people can trace to solve problems for which a solution consists of discrete, deterministic, unambiguous atomic operations [14]. But ML initiatives differ by nature from "traditional CT" with regards to the role of algorithmic steps. For one, it is practically impossible to trace how a neural network reaches its solution, and the individual "steps" in the model are not central. Many applications hide most of what happens with the individual neural network operations. What is visible to the user in ML education tools can range from chiefly algorithmic [112], [113] to chiefly data-driven [76], [119], [120].

A number of ML education tools do not require exposure to the concept of algorithmic steps—on the contrary, some celebrate the shift from "coding to teachable machines" [15]. In training a model, in testing a model, and in connecting the model with actions, there is less a need for thinking about algorithms, and more concern with describing users' intentions, and especially with designing data sets that can be used for creating the desired behaviors [Authors2020a,2020b]. Learning about ML systems in the classroom does not need to be predominantly about traceable, procedural execution of operations but about getting enough of the right kind of data for the job [54], [99]. Learners need to understand how ML "reasons" using models and representations built from data, and not rules coded in the system [54], [99]. Designing an algorithm gives way to derivation of models operating at a meta-level. That does not, however, diminish the need for teaching traditional CT, because real apps are almost always a combination of data-driven and rule-driven computing. ML is an addition, not a replacement.

### 7. Glass boxes and black boxes are relocated.

All educational approaches in computing education contain black boxes. The choice of where those black boxes are located is essentially a choice of what children can explore with the system, and what will they not learn about [72]. Traditional computer programs are glass boxes to the point where their flow of program execution, changes in values of variables, and everything else a program does are all hand-coded in the program, and the program flow can be tracked, visualized, and paused at any point to examine the program state at any given step of execution [85]. Most ML models are more opaque in the sense that the weights and parameters of neural networks and regression models are not set by hand, but they are trained by feeding the model a lot of data, each sample adjusting the model's internal parameters bit by bit. From the viewpoint of traditional programming, the weights and parameters of a trained model are a black box in the sense that examining them lends no interpretation on what the neural network does or how. Many AI in K–12 education initiatives consider the public perception of AI as black boxes a major challenge for AI education [35], [41], [54], which makes explainable AI (XAI) a major opportunity also in K–12, with example applications emerging [Authors2020e].

The location of black boxes shift over learners' skill progression. In early machine learning education children can train ML models and use them to control the computer (for instance, with Wolfram Programming Lab or Snap!), but the actual mechanisms remain a black box, and can only be opened much later in the school. In research on ML education the opaqueness of ML models has faced issues known to K–12 educational technology designers through ages [35], [72] [Authors2020b,2020e]: The more accessible the systems are for novice learners, the more actual ML mechanisms they need to hide from the users—and the less students learn about what goes on under the hood. However, experiments with older, high school students building an object recognition system from the scratch have showed that ML models that rely on very few features can be programmed from scratch in elective programming classes [Authors2019a].

### 8. The notional machines evoked in machine learning differ greatly from those of traditional programming.

As Sorva [86] wrote, much of computing education research has focused on overcoming the challenge of how to develop a robust conception of a notional machine (a mental model of what the computer does when it executes a piece of code). In the extensive body of literature on notional machines with novices and experts [86], little is relevant to ML. For instance, neural networks and regression models function very differently from traditional programs both at the abstract level and at the concrete level: In the abstract sense, passing data through a neural network is conceptually very different from the execution flow of a traditional program. In the practical sense, neural networks are embarrassingly trivial to parallelize on thousands of special-purpose processing cores that any cheap graphics card has [49]. But the more the ML processes are black-boxed, the more limited the prospects of constructing accurate notional machines related to them [35]. The notional machines and mental models for ML in computing education are bound to be very different from those of traditional programming education (Authors2021), and there is little research on the former in computing education research body of literature.

### 9. Machine learning systems are tested and debugged differently from rule-based programs.

Earlier research on debugging strategies in K–12 computing education offers little advise for those ML education systems where children can train their own ML models. That is because "debugging" of ML models turns out to be very different from debugging traditional rule-based programs [35], [120]. Many machine learning algorithms are "soft" in the sense that their results are not straightforwardly discrete but tell, for instance, the probabilities of the input belonging to different classes—for instance, of Google's K–12 TM2 tool shows, as percentage, the model's confidence in classifying an image. Machine learning models are typically also

"brittle" in the sense that when trained in one environment with one kind of data, small changes in the environment or input data may render the model nearly useless. Children have learned the concepts of softness and brittleness, for instance, when trying their mood recognition system in a different location with different background [Authors2020a]. That issue has been widely recognized in ML education, and there is research on using very simple examples to teach high-school students concepts like overfitting, accuracy, precision, and recall [22].

What is more, the workflows and philosophy of testing, debugging, and fixing ML models differ greatly from debugging traditional programs. Debugging traditional program code should not be based on trial and error, but on systematically tracking and understanding what happens on each line of the code. On the contrary, training black boxed ML models is ultimately based on trial-and-error type search of optimal hyperparameter and feature space. A number of initiatives have focused explicitly on teaching ML workflows, described in different ways but typically including variants of "data collection, data entry, data visualization, feature engineering, model building, model testing, and data permissions" [88] [Authors2020d], depending on the model.

### 10. Judging system goodness becomes more complex.

Machine learning shifts the attribute of "goodness" from correctness to effectiveness. The main epistemological stance of traditional, rule-based computing used to be correctness and verifiability in deterministic systems [56], which then expanded to more holistic views of computing systems, such as the DRUSS principles (dependable, reliable, usable, safe, and secure). In many cases ML solutions can at best be "probably approximately correct," their goodness statistically determined [102], which is reflected in many ML in K–12 initiatives [22], [36] [Authors2018]. Effectiveness has, of course, been a long time keyword in K–12 computing education—take, for example, RoboCup football matches between children's robots, where no system can be "correct" but goodness of systems is measured by how often they win against other football-playing robots. It turns out that teaching learners about the strengths and weaknesses of ML-based technology can differ greatly from the same related to rule-based programs [54].

Reductionism, or the ability to fully explain the functioning of a large system by the functioning of its basic, atomic elements, gives in many ML applications way to emergence, where complex systems have properties that only arise from the interactions of large numbers of interacting parts. On a continuum from a bivalent view of correctness (the program outputs are either correct or incorrect) to a more fuzzy view that involves reliability, efficiency, and the like, ML is situated more towards the same end with educational robotics and other tangible learning environments, that rely on contextual, relative, and pragmatic view of goodness.

**11. Approaches to STEM/STEAM integration change.**

Computing education has for decades been gradually moving from closed problems towards open problem classes along many dimensions [89]. Computing in STEM education has increasingly embraced open-ended, authentic, and hands-on education, too [8]. A number of existing ML tools have the potential to further facilitate that shift in the domain and nature of learning problems and projects in the STE(A)M classroom [15], [33], [62]. ML projects at schools have fruitfully started new approaches to complement earlier STEAM integration approaches (traditionally focusing on, for instance, the principles of procedural or reactive programming with sensor-based artefacts [33], [46], [112] [Authors2020e]).

In the STEAM class, students can first explore what ML systems are capable of—for example, they can experiment on training ML models on graphical user interface based systems or drag-and-drop systems [46], [112]. Once familiar with how ML works, students can start by defining the problem, issue, or concern in STEAM fields they want to solve. This squares very well with the changing learning process when moving from STEM to STEAM education. STEAM education emphasizes real-world problems and exploration of multiple solutions to them, and ML is at its best with rich, real-world data. A number of K–12 ML platforms designed for rich multimodal material (pictures, sound, movement tracking) fit well STEAM education in those cases where laws, formulas, and rules may not be readily available. Example data sets used in K–12 ML education include data from the bicycle sharing system in Chicago, Spotify song feature data set, and face recognition data sets [61], language corpora from Dr. Seuss, Taylor Swift, and Shakespeare [105], and there is an example of teaching ML concepts in the middle school using the advanced RapidMiner software, making models of mango sweetness, mango quality, and the mango market [76]. AI has been taught along with the philosophy and history of science, too [33], [84]. Many ML-based learning environments offer high degrees of freedom for a broad variety of experiments.

**12. ML helps students learn how their world works.**

One tenet of technology education and computing for all initiatives has been to teach students how the world around them works [31]. Computing education should be able to banish magic from computing systems by exposing the mechanisms by which they work. Educational technology efforts from the Little Man Computer to block-based programming have done a great job teaching how rule-based systems work [32], [96]. But many features of services most familiar to children today—such as TikTok, Spotify, Youtube, and Netflix—as well as many features of technology children use daily—such as face recognition, speech recognition, recommenders, and targeted advertisement—are better taught using ML based educational technology than they are with traditional programming [Authors2019c]. Many ML in K–12 initiatives explicitly address the need to teach children the aspects of modern AI that affect their life and future work [10], [15],

[35] [Authors2020d]. Again, data-driven and rule-driven approaches both need to be taught, and ML does not make teaching rule-based programming obsolete. Yet, ignoring ML in computing education leaves open a gap that is widening by the day.

**13. Applications of ML technology bring about some new ethical concerns to be included in computing education.**

ML in K–12 educators have argued that any instruction on ML needs to also include ethics of AI [99] and challenge the mythical AI narratives from the popular media and Hollywood [3], [21], [54]. It has been suggested that AI needs to be demystified [37], and magic banished from it [41]. Many initiatives take the ethical challenge very seriously: For instance, the extensive German AI/ML learning package deployed to thousands of schools had a heavy focus on ethical and societal aspects, including scenario work for what kind of a future would we want from learning machines [78]. What is more, ML classes aimed at high school girls have been used to encourage broader participation in computing [101]. The rationales for including ethics differ only to a limited degree from the rationales for establishing computing ethics as a cross-cutting computing curriculum subject in the 1990s [94]. Yet, ML does bring new perspectives to topics like privacy, surveillance, job losses, misinformation, diversity, algorithmic bias, transferability, and accountability, among others [54].

## V. PITFALLS AND WEAKNESSES

MACHINE LEARNING provides to computing and automation a perspective that is markedly different from the perspective of rule-based computing and programming-based computational thinking. It reaches to some areas that coding initiatives cannot reach, while the opposite is equally true. ML in K–12 literature lists a large number of weaknesses, pitfalls, and dangers related to ML in K–12 education.

Firstly, especially in the lower school grades and with "low floor" apps, learning can be shallow and superficial, because children learn workflows, not the internal ML mechanisms. Those workflows are described differently in different initiatives. For example, one initiative described their workflow as data collection, data entry, data visualization, feature engineering, model building, model testing, and data permissions [88]. Another described it as requirements analysis, collecting the training data sets, training the model, evaluating the model, and deploying the model to work in apps [Authors2020a]. Also parts of the workflow can be black-boxed: One study focused on just two stages of the workflow, data labeling and evaluation, and black-boxed the other parts of the workflow [35], [36].

Secondly, there is no consensus over the trade-offs necessitated by black-boxes, and there is a paucity of research on their effects. Interacting with highly opaque processes may lead children to develop inaccurate or oversimplified notional machines, which can be difficult to change once formed [35]. Black-boxing also risks unrealistic expectations of what ML systems can do, and in case of younger children, anthropomorphizing and personification of ML systems [17], [33], [54], [108]. At the same time, ML in K–12, especially with younger learners, requires hiding much of the underlying complexity. Research on explainable AI in K–12 is forming up [Authors2020e], and there are initiatives to expose the implementation mechanisms of AI in high school classrooms [41], [58], but the question of black-boxing in K–12 ML education remains essentially an open question [35].

Thirdly, compared to ML systems, classical AI systems are better for learning the rules underlying traditional AI (because children program the rules in them)—take, for instance, expert systems or chatbots [114]. Rule-based programming is based on relatively simple Boolean logic and discrete mathematics, the basics of which can be taught relatively early, and then deepened at age-appropriate levels from primary school to high school. By contrast, ML is based on advanced mathematical concepts beyond the children's grasp, such as statistics and probabilistic modeling [21]. What is more, opening the ML black box in education requires education to address skills from both extremes of the applied–formal spectrum: very practical and very theoretical [21]. ML in K–12 initiatives share the challenge of how to ease the entry to ML and avoid advanced mathematics [21], [22].

Fourthly, there is no agreement over the relationship between ML skills and knowledge and the multitude of skills and knowledge labeled "computational thinking" (CT) [95]. As nearly all ML apps are really combinations of rule-based code and ML models, it has been suggested that traditional CT frameworks should be extended with ML-related concepts, such as classification, prediction, and generation [104], [105]. At a more general level, it has been argued that if computational thinking is about how to make computers do jobs for people, ML is, by definition, a part of computing's disciplinary ways of thinking and practicing—in other words, a part of computational thinking [14].

Fifthly, compared to computing education research on programming in K–12, and even compared to research on teaching rule-based "good old-fashioned AI" in K–12, research on how to teach ML in K–12 is in its infancy [17], [35], [54]. Lack of experience on age-appropriateness of ML concepts has caused surprises, such as cases where K–6 children grasp seemingly much more sophisticated ideas quickly but simpler-looking ideas may take longer [33]. In the absence of long-term experience from ML curriculum implementations at different school levels, continuity of AI/ML education from kindergarten to high school is an equally daunting challenge as it is in traditional computing education [33]. And the pet peeve of CT critics—"How does one measure student progress in CT?"—applies equally well to ML, too.

Finally, a topic high on the hype cycle risks unrealistic expectations and misconceptions. ML education in K–12 should not repeat the inflated expectations about Logo that ultimately led to disappointment when the initiative could not fully deliver to its promise [4]. Although ML makes possible automation of new classes of jobs that could not be automated before, it is still very limited in its applications, and much of the ML hype in the popular press is based on limited understanding of what ML actually does [11]. Black boxes further obscure functionality, and the "Eliza effect"—where a system appears to observers as much smarter than what its internal functioning warrants [54]—is a real danger in ML education.

## VI. CONCLUSIONS

THE FRONTIERS of computing education have been in constant flux throughout the disciplinary history of computing [96]. Computing education has always followed the state-of-the-art technology in the field, with the help of improved interfaces and layered stacks of abstraction levels to hide the underlying complexity. As machine learning is making its way to becoming mainstream technology [49] and core computing knowledge [81], [82], and as ML applications have become commonplace [9], computing educators have anticipated a shift in K–12 computing education, too [15].

The nascent body of literature on teaching ML in K–12 education is rapidly developing along a route that is in a number of ways different from the traditional programming and computational thinking oriented computing curricula. That new kind of computing education is characterized by a number of opportunities and changes in thinking, including:

1) New classes of real-world applications for classroom experiments
2) Shift from rule-driven to data-driven thinking
3) Change in the role of syntax and semantics
4) Activities well aligned with modern pedagogy
5) Access to bodily and natural language interaction
6) A shift away from algorithmic steps
7) Higher level of abstraction and black-boxed mechanisms
8) A need for new notional machines
9) New models of testing and debugging
10) New attributes of goodness of programs
11) Deeper integration with STEAM subjects
12) Ability to explain many services children use daily
13) Direct connections to topical issues in AI ethics

But as schools and teachers struggle with the recent wave of integrating computational thinking into school curricula [13], teaching ML in K–12 poses an even more daunting challenge. The computing education research body of literature contains remarkably little research on how people learn to train, test, improve, and deploy ML systems [15], [81], [82]. Research on ML education is in its infancy, most reports being either analytical—concept development, essays, or curriculum descriptions—or, if empirical, of the exploratory, proof-of-concept, or Marco Polo types ("I went there and I saw this"). Literature reviews have only recently started to emerge [59], [117].

As it has taken computing educators many decades to bring traditional programming and computational thinking into school curricula in many countries [31], one can question whether time is ripe yet for expanding computing education with another addition. Schoolteachers are already struggling with the recent wave of integrating computational thinking into school curricula [13], [57]. Computing, computational thinking, or programming, is restricted to a rather limited number of classroom hours in national curricula; or have been more, or less, successfully "integrated" into other subjects. Perhaps that time is better spent on the better researched traditional computing curricula than a newer topic on which there is little literature, material, classroom examples, readily available learning objects, or pedagogical experience.

But despite the challenges, there is a clear need for understanding how ML-based and data-driven systems in people's everyday lives work [34]. Thus, there is a need to build ML education from ground up: research is needed for pedagogical models, skill progression schemes, appropriate educational technology, ethical dilemmas, domain integration, and all other elements of education. This article has outlined some elements of education that computing educators need to consider regarding machine learning in the K–12 classroom. This is a vital area, if future citizens are to be empowered in regard to the systems around them, and that warrants considerably more focus from the computing education research community.

...